\documentclass[12pt,a4paper,prd,eqsecnum,preprint]{article}
\usepackage{subeqn}
\usepackage{graphicx}
\usepackage[footnotesize,bf]{caption} 

\begin{document}
\title{ Effects of Pair Creation on  Charged Gravitational
         Collapse} 

\author{Evgeny Sorkin\thanks{email: sorkin@merger.fiz.huji.ac.il}
  \ and Tsvi Piran\thanks{email: tsvi@nikki.fiz.huji.ac.il} \\ 
  \small 
  \textit{
    The Racah Institute of Physics, The Hebrew University, 
    Jerusalem, Israel, 91904
    }
  } 
\date{}
\maketitle

\begin{abstract}
  We investigate the effects of pair creation on the internal geometry
  of a black hole, which forms during the gravitational collapse of a
  charged massless scalar field. Classically, strong central
  Schwarzschild-like singularity forms, and a null, weak, mass-inflation
  singularity arises along the Cauchy horizon, in such a collapse.  We
  consider here the discharge, due to pair creation, below the event
  horizon and its influence on the {\it dynamical formation } of the Cauchy
  horizon.  Within the framework of a simple model we are able to
  trace numerically the collapse. We find that a part of the Cauchy
  horizon is replaced by the strong space-like central singularity.
  This fraction depends on the value of the critical electric field,
  $E_{\rm cr}$, for the pair creation.
\end{abstract}
\section{Introduction}
\label{intro}
The well known exact solution of the coupled Maxwell-Einstein
equations outside the spherically-symmetric matter distribution is the
Reissner-Nordstr{\o}m solution. The analytic extension of the
Reissner-Nordstr{\o}m metric has rather exotic properties.  The black
hole's interior contains Cauchy horizons, time-like singularities and
tunnels to other asymptotically flat regions.

\begin{figure}
\centering
\noindent
\includegraphics[width=10cm]{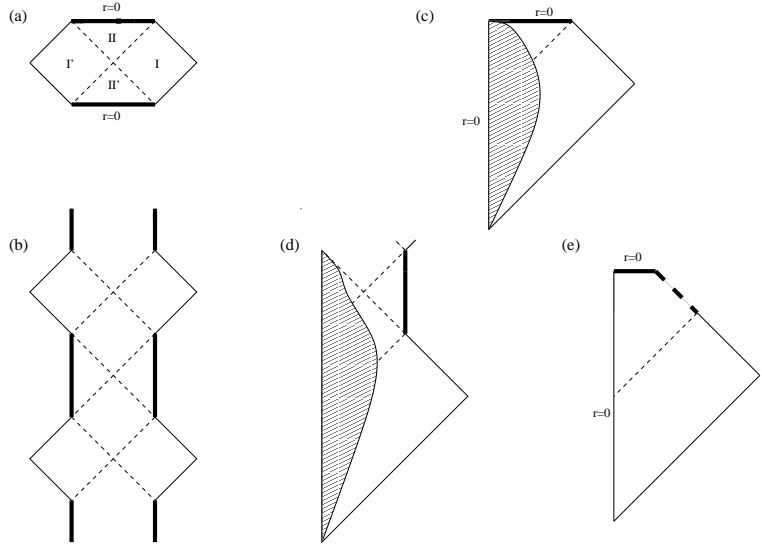}
\caption[Penrose diagrams of Schwarzschild, Reissner-Nordstr{\o}m
and dynamical space-times] { Penrose diagrams. $(a)$ and $(b)$ are,
  respectively, the Schwarzschild and the Reissner-Nordstr{\o}m
  eternal (pre-existing) spacetimes , $(c)$ and $(d)$ are ones
  expected to form during a dynamical collapse of a neutral and a
  charged matter, respectively.  The actual, dynamically calculated,
  space-time geometry for a charged gravitational collapse, $(e)$ is
  dramatically different from that depicted on $(d)$ and it resembles
  (in broad terms) the Schwarzschild space-time.  In all the figures
  thick solid lines represent central singularities (space-like for
  the Schwarzschild manifold or time-like for the
  Reissner-Nordstr{\o}m one). The weak, null singularity in figure
  $(e)$ is represented as a thick dashed line, while thin dashed lines
  describe various horizons.}
\label{fig:Penrose}
\end{figure}

Recently, it has been shown both in perturbative analysis and by
solving the full non-linear problem that a Cauchy horizon inside a
charged black hole is transformed into a null, weak singularity
\cite{His,IsPo,Ori1,Ori2,BrSm,Bur}.  The Cauchy horizon singularity is
weak in the sense that an infalling observer crossing it experiences
only a finite tidal deformation \cite{Ori1,Ori2}.  However, the
curvature scalars diverge along the Cauchy horizon, leading to an
unbound growth of the internal mass-parameter, a phenomena known as
the mass-inflation \cite{IsPo}. The earlier studies were done on the
pre-existing (eternal) Reissner-Nordstr{\o}m space-time.  Hod and
Piran \cite{HodPir}, have demonstrated explicitly that mass-inflation
takes place also during a dynamical charged gravitational collapse.
The dynamical space-time is drastically different from the
analytically extended Reissner-Nordstr{\o}m manifold: it resembles
more the Schwarzschild one. The Penrose diagrams of the various
space-times are depicted on Figure \ref{fig:Penrose}.

This is so far the classical picture.  Our goal, here, is to consider
quantum effects and to investigate the influence of pair creation in
strong electric fields on charged gravitational collapse. Specifically,
we are interested in the effects of pair creation on the inner
structure of the black hole that forms in such a collapse. This goal
differs from previous works on the subject.  Pair creation was mainly
considered in the external region of a pre-existing black hole's
space-time, outside the event horizon
\cite{ZCGDR1,ZCGDR2,ZCGDR3,ZCGDR4}. It has been shown that the
produced particles rapidly diminish the charge of a black hole as seen
by an external observer.

Particles creation takes place, however, also in the inner region of a
black hole.  Novikov \& Starobinski\u{i} \cite{NovStar} and Herman \&
Hiscock \cite{HerHis} studied the inner geometry and the stability of
a Cauchy horizon in the pre-existing Reissner-Nordstr{\o}m space-time
influenced by the pair creation effect. The model in Ref. \cite{HerHis}
assumes the instantaneous disappearance of the electric field, when
the pair creation takes place inside the event horizon along $r_{\rm
  cr}=const$ hypersurface. The Reissner-Nordstr{\o}m patch of the
space-time exterior to this hypersurface is glued along the $r_{\rm
  cr}=const$ hypersurface to an interior Schwarzschild patch. In this
model the Cauchy horizon  does not exist.  The model in Ref.
\cite{NovStar} assumes an initially Reissner-Nordstr{\o}m geometry and
allows the evolution of the electric field through the back-reaction
of the created pairs. In this model the initial Reissner-Nordstr{\o}m
geometry evolves to an uncharged, Schwarzschild-like one - the Cauchy
horizon  is shown to be unstable with respect to the process of pair
creation. It should be emphasized that in this model the Cauchy
horizon  is assumed to be stationary $r_{\rm CH}=const$
hypersurface. This is in contrast to the recent investigations
\cite{His,IsPo,Ori1,Ori2,BrSm,Bur,HodPir}, where the Cauchy horizon
was shown to be non static, contracting null hypersurface.

Although, the particle production in
the intensive electric field is the simplest quantum process, its
influence on the inner structure of black holes was not studied yet in
an evolutionary context.  The effect of  pair creation in the
intensive electric field is probably most important, when dealing with
a formation of the inner structure of a charged black holes.  This
depends, of course, on the parameters of the formed black hole.  In
this work we take a different point of view from \cite{NovStar,HerHis}
and explore the dynamical picture i.e., we replace the question is the
Cauchy horizon stable, with the question does it form at all.  To
address this question, one should consider the collapse of a charged
self-gravitating matter.  The electron-positron pairs are produced in
the electric field of the collapsing matter. To treat consistently the
problem of such a collapse, one should take into account the
back-reaction of the produced pairs on the source's electric field.
This is achieved by adding the electric current due to the produced
pairs as a source to the Maxwell equations:
\begin{equation}
\label{maxInt}
{F^{\alpha\beta}}_{;\beta}=4 \pi J^{\alpha}_{\rm free} +4 \pi J^{\alpha}_{\rm
    pairs}  \ .
\end{equation}
The charge conservation equation in the situation with  pair
creation is modified by adding the charge source:
\begin{equation}
\label{chargeoons}
{J^{\alpha}}_{;\alpha} = \Gamma(F^{\alpha\beta}) \ .
\end{equation}
 The stress-energy of the electric current of the produced particles arises
from the stress-energy of the electric field. The latter is the source of
the pairs, by means of the energy-momentum conservation.

To formulate the problem properly we need a back-reaction formalism
that includes the back reaction of the pairs on the stress energy
tensor of the electric field that create them. Without this the
problem would not be self consistent.  Such a formalism is not
available.  Instead we consider here a toy model that utilizes the main
physical properties of the system - the fact that the pairs limit the
electric field to a critical value, $E_{\rm cr}$.  We describe this
effect of pair creation by introducing a nonlinear dielectric constant
which prevents the electric field from exceeding $E_{\rm cr}$, the
critical pair creating field. In doing so we have ignored the electric
current of the pairs and their stress-energy.  We also disregard the
contribution from vacuum polarization, which becomes significant only
for the exponentially large fields \cite{GMM}.  In spite of this
simplifying assumptions we believe that this model captures the
characteristic behavior of the real system.

In section \ref{motiv} we consider discharge in a classical
space-time and give the motivation of the investigation of the influence of
the pair creation on the inner structure of charged black holes.  In
section \ref{model} we present the underling physical model. We
develop the formalism and discuss the applicability of the model.
Section \ref{numerics} describes our numerical scheme.  The results
are presented in section \ref{results}. We compare a classical charged
gravitational dynamical collapse with a collapse with the discharge.
We summarize our conclusions in section \ref{Summary}.  We use units
in which $c=G=\hbar=1$.
\section{Discharge in a  Classical Charged Space-time}
\label{motiv}
When pairs are created in an asymptotically flat region
one of the particles, having the same charge as the field's source, is
repulsed from the body and escape to infinity. The another member of
the pair is attracted to the body, decreasing its charge.
This occurs, for example, to pairs created in the field of a charged black
hole, outside its event horizon. The black hole discharges rather
quickly, until its external field becomes subcritical
\cite{ZCGDR1,ZCGDR2,ZCGDR3,ZCGDR4}.

\begin{figure}
\centering
\noindent
\includegraphics[width=10cm]{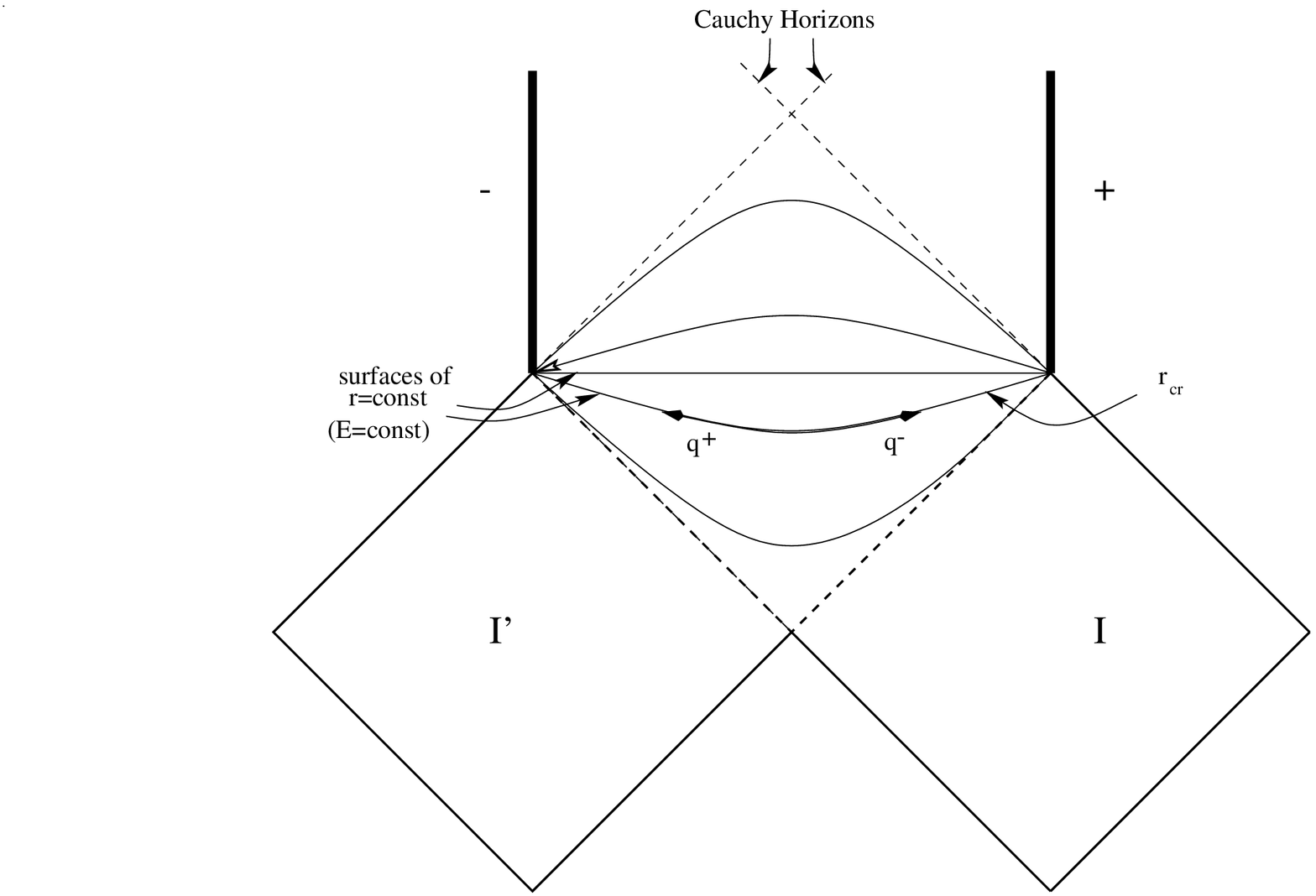}
\caption[Pair creation below an event horizon]{
  Pair creation inside the event horizon.  Here we depict the Penrose
  diagram of the Reissner-Nordstr{\o}m space-time. Pairs are produced when the electric
  field reaches the critical value $E_{\rm cr}$ along the surface $r=r_{\rm
    cr}$.  Oppositely charged particles are accelerated in opposite
  directions, leading to the redistribution of the charge bellow the
  event horizon and to a change of the inner geometry of the charged
  black hole.}
\label{fig:qed}
\end{figure}

A more interesting situation occurs when a significant pair creation
takes place within the event horizon of a charged black hole.  The
newborn particles do not have a spatial infinity to escape to, they
are trapped within the event horizon. In the Reissner-Nordstr{\o}m
manifold in the region between the outer and the inner horizons the
area coordinate $r$ and the time change their roles.  The vector
$\partial /\partial r$ is now timelike, while the vector $\partial
/\partial t$ is spacelike. The only non-vanishing components of the
electromagnetic tensor are $F_{uv}=-F_{vu}$ and only $F_{rt}=-F_{tr}$
are nonzero.  An infalling observer moving along $d t/ d\tau = 0$
world-line, in the region between the horizons, will experience a
spatially homogeneous electric field, increasing in strength into the
future (the $r$-coordinate decreases). The electric field has only a
$\hat t$-component: $ {\bf E} = {q(r,t) \over r^2} {\partial \over
  \partial t}$, wherein the bold face denotes usual 3-vectors. The
direction of the field lines in a regular Reissner-Nordstr{\o}m
space-time is from one singularity to the other (see Figure
\ref{fig:qed}).  The maximally extended Reissner-Nordstr{\o}m
space-time has a charge asymmetry in the sense that two external
observers in the two past asymptotically flat regions, $I$ and $I'$,
see the black hole charged oppositely. The left-hand and the
right-hand singularities seem to such observers to have opposite
charges.

In the interior of a classical charged black hole, between the inner
and the outer horuzons, pairs of charged particles are produced by the
electric field along $r=const$ surfaces.  An oppositely charged
particles are accelerated in the opposite $\pm\partial /\partial t$
directions.  Thus, if the black hole has a negative charge the
positively charged particles will be attracted to it, accelerating toward
the left-hand singularity, while the negatively charged particles will
be repulsed from this singularity, accelerating toward the right-hand
one (see Figure \ref{fig:qed}).

This leads to the redistribution of the charge, which was initially
concentrated near the left-hand singularity. At the end of this
process, in the perfectly symmetric situation, when the charge is
equally divided among the left-hand and the right-hand singularities
the electric field disappears. But the whole exotic structure of the
inner part of the analytically extended Reissner-Nordstr{\o}m manifold is due to the
existence of this electric field. Therefore, vanishing of the electric
field leads to the disappearance of the tunnels to other
asymptotically flat regions. We will show here that a
similar redistribution of the charge takes place also in a
dynamical space-time, leading, as we shall see, to the partial closure
of the ``space tunnels''.

\section{The Physical Model}
\label{model}
In this section we develop an evolutionary formulation
which includes the effect of pair creation  in  strong electric fields.
We present bellow a simplified toy model, describing this effect
for a dynamical space-time in which a  black hole forms.
\subsection{The Formulation}
\label{ClassEM}
The first study of  charged particles production in an uniform electric
field was undertaken by J.Schwinger in 1951.
The Schwinger's formula for the number of scalar pairs created by the
field $E$ per unit four-volume is:
\begin{equation}
\label{Schwingerrate}
{\Gamma = {e^2 \over{2\pi^2 \hbar^2 c}}E^2\sum_{n=1}^{\infty} {(-1)^{n+1} \over {n^2}}
    \exp({- n E_{\rm cr} \over E})} \ ,
\end{equation}
where  $m$ and $e$ are the mass and the charge of a created particle
and the {\it critical electric field}, $E_{\rm cr}$, is defined as:
 \begin{equation}
\label{CriticalField}
{ E_{\rm cr} = {\pi m^2 c^3\over {e \hbar}}} \ .
\end{equation}
The  production rate of  fermions  differs from
(\ref{Schwingerrate}) by an overall factor $2$ and by the absence  of the sign
interchange $(-1)^{n+1}$.

The effect of the vacuum polarization (the change of the vacuum
electric permittivity) in strong electric fields is instantaneously
stronger than the contribution from a pair production ( by $
\ln(E/E_{\rm cr}) $ for $E \gg E_{\rm cr}$), but the latter can
accumulate with the time  \cite{GMM}. The
integrated contribution from the pair creation can dominate the
contribution from the vacuum polarization.

Our model utilizes the essential property of the  Schwinger's result
(\ref{Schwingerrate}) - the exponential dependence on the ratio $E_{\rm cr}/E$,
which means that pair creation rate is exponentially large for a
supercritical field and the rate is exponentially suppressed for a
subcritical field.  This dependence suggests that once the field rises
above $E_{\rm cr}$, charged pairs are produced intensively, reducing the field
down to the critical value.

We neglect $(i)$ the net electric current of the pairs and $(ii)$ the
stress-energy of the produced particles, assuming that the role of the electric
current is confined to prevent the electric field from rising above
the critical value.  Hence, the electric field
will be taken as:
\begin{equation}
\label{ModelField}
E\equiv
 \left\{
   \begin{array}{r@{\quad \rm{if} \quad}l}
     E_{\rm ordinary} & E_{\rm ordinary}<E_{\rm cr} \\ E_{\rm cr} & E_{\rm ordinary}\ge E_{\rm cr}
   \end{array}
 \right. \ .
\end{equation}
The $ E_{\rm ordinary}$ stands for the ordinary electric field which
would have arised in the absence of pair creation. 

In this case we can  mimic the effect of
particles production as if the system was placed in a dielectric medium.
Effectively, the polarization of this medium prevents the electric
field $E$ from rising above $E_{\rm cr}$, while the {\it electric
  displacement}, ${\bf D}={\bf\epsilon}{\bf E}$, changes.
  The scalar quantity ${\bf\epsilon}$ is the {\it dielectric
  constant}.
The electric displacement is related to the density of a free charge
via the Maxwell equation:
\begin{equation}
\label{d=v^2(ro)}
\nabla\cdot{\bf D}=4\pi\rho_{\rm free}  \ .
\end{equation}
The dielectric ``constant'' that leads to
(\ref{ModelField}) is given by:
\begin{equation}
\label{def_eps}
\epsilon \equiv \left\{
  \begin{array}{r@{\quad \rm{if} \quad}l}
     1 & E_{\rm ordinary}<E_{\rm cr} \\ |{\bf D}|/E_{\rm cr} & E_{\rm ordinary}\ge E_{\rm cr}
   \end{array}
 \right. \ .
\end{equation}

The local description of the classical theory of electromagnetism in a
curved space-time in a dielectric media can  be derived from an effective
local Lagrangian:
\begin{equation}
\label{effLag1}
{\cal L}^{\rm{(eff)}} = -{1\over 8\pi}{\bf E}\cdot{\bf D}
= -{1\over 8\pi}{{|{\bf D}|}^2\over\epsilon} \ .
\end{equation}
We use this Lagrangian with $\epsilon$ given by (\ref{def_eps}) to
describe the effects of  pair creation in the strong electric
field.

Our toy model captures the essence of the physics, particularly 
the fast reduction of the supercritical field down to the critical
value, $E_{\rm cr}$ and the energy-momentum conservation. The dielectric
``constant'' that we introduce has the same effect as the pairs which
``shorten'' an electric field above $E_{cr}$.  However, we ignore all
other features of the pairs, specifically we ignore the pairs
themselves, their energy momentum tensor (which is replaced by a
modified electromagnetic energy-momentum tensor that arises from the
dielectric constant) and their electric current.  This is done in
order to obtain a simple self consistent energy conserving system.  It
is difficult to estimate what would be the effect of the electric
current of the pairs. On the other hand it is clear that the
stress-energy of the pairs would make a positive contribution to the
mass-parameter (\ref{massfunc}) and will make the effects, which we
describe latter, more pronounced.

Another artificial feature of our model is the ad hoc introduction of
$E_{cr}$. This critical field is associated with the mass of the
charged particles (see Eq. (\ref{CriticalField})). However, for
simplicity our model is based on a massless charged scalar field, whose
characteristics are along null geodesics. For such a massless field
the critical electric field vanishes and charged massless pairs
are produced even for an infinitely small electric field.  We
introduce a critical field $E_{\rm cr}$ as a {\it free parameter} which can be
used to define the mass of the created particles trough the relation
(\ref{CriticalField}).

In addition to the above properties there are  other minor and
physically  justified  assumptions:
First, our toy model ignores the contribution from vacuum
polarization due to the intensive electric field (to be distinguished
from the effective polarization, which we describe here). This would be
justified if the effect of the vacuum polarization is small compared
to the pair creation contribution. This is in fact the situation when
the electric field is not exponentially large \cite{GMM}.

Second, when constructing our model, we have utilized Schwinger's
result (\ref{Schwingerrate}), more precisely, its exponential
dependence on $E_{\rm cr}/E$. Schwinger's formula is valid, strictly
speaking, only for an uniform and static electric field over a flat
space-time background.  The approximation of flat-space is valid if
the radius of curvature of the dynamical geometry is much greater than
the Compton wavelength of a created particles.  The radius of a
curvature can be taken of order $R^{-1/2}$, where $R$ is the Ricci
curvature scalar. The Compton wavelength of the particles is
$\ell={\left({\pi c \hbar \over e E_{\rm cr}}\right)}^{1/2}$. Thus, the
condition for a flat-space approximation, $\ell \ll \ R^{-1/2} $,
takes the form
\begin{equation}
\label{cond1}
{\left({{\pi c \hbar \over e E_{\rm cr} } R}\right)}^{1/2} \ll 1 \ .
\end{equation}
But, the Ricci scalar $R$ has been shown, in dynamical models, to
diverge approaching the Cauchy horizon, (this is the mass inflation scenario) thus,
the approximation breaks down in vicinity of the singular Cauchy horizon.
Notwithstanding, our model is, actually, not based on exact Schwinger
result, but on its exponential dependence on $E_{\rm cr}/E$ which is
non-perturbative.

Anyway, all approximations will be broken at some moment.  Eventually,
the curvature in the vicinity of the singular Cauchy horizon \ becomes Planckian
since the ``Coulomb component'' of the Weyl curvature diverges
exponentially with advanced time (for a spherical symmetry- $|\Psi_2|
\sim m/r^3$, with $m$ the internal mass parameter.)  Moreover, the
Ricci curvature may dominate the Weyl curvature and surpass the Planck
values even earlier.  In either case our analysis becomes meaningless,
and a theory of quantum gravity is needed.  We obviously consider only
the sub-Planckian regions.

\subsection{The Equations}
Our goal is to integrate numerically the evolution equations and
to follow the collapse of a spherically symmetric regular initial
scalar field distribution via the formation of an apparent horizon and
a Cauchy horizon, toward a central singularity. The conventional choice of
coordinates for this dynamical evolution is  double-null coordinates.
In these coordinates: $(1)$ The apparent horizon (when it forms) is
regular i.e. it is free from unphysical coordinate singularities;
$(2)$ For a massless scalar field the characteristics are null so this
choice is ``natural'' for models involving massless scalar fields.

We  choose the line element of the form:
\begin{equation}
\label{metric}
{{ds}^2=-{{\alpha(u,v)}^2 du dv} +{r(u,v)}^2 d{\Omega}^2     }  \ ,
\end{equation}
where  $d{\Omega}^2$ is the unit two-sphere.
There is a coordinate gauge freedom: the choice of coordinates ${u,v}$
is unique only  up to a change of variables $ v'=f_1(v), u'=f_2(u) $,
which leave the line element (\ref{metric}) unchanged.  For the time
being we do not specify our double-null coordinates: these are just
general ingoing and outgoing null coordinates. Later, when discussing
the numerical integration we will fix the gauge freedom and specify
the coordinates.

Let the $F_{\mu\nu}$ be the electromagnetic tensor, defined as
$F_{\mu\nu}\equiv A_{[\nu;\mu]}$.
In a spherically symmetric space-time the only non vanishing field
components are: $F_{uv}=-F_{vu}$. Thus, only $A_ v$ and $A_u$ need be
non vanishing.  Then $A_v$ can be removed by the gauge transformation
$A_{\alpha} \rightarrow A_{\alpha}+\Lambda_{;\alpha}$ , with $\Lambda=
-\int A_v dv$. We are left with $A_u \neq 0$ and we denote:
$a(u,v)\equiv A_u$.

We reformulate the set of coupled Einstein-Maxwell-Scalar Field
equations as a first order system. The numerical integration of the
first order system functions very well both for the uncharged case,
see Ref. \cite{HamSt}, and for the charged situation
\cite{HodPir}.  It is convenient to define the auxiliary variables:
\begin{equation}
\label{newvar}
d\equiv{\alpha_v\over\alpha},  \  f\equiv r_u,  \    g\equiv r_v,\    s\equiv\sqrt{4\pi}\psi,\
  w\equiv s_u,\     z\equiv s_v \ ,
\end{equation}
wherein $\psi$ is the complex massless scalar field.  We have adopted the
notation $W_x \equiv {\partial W / \partial x} $ for partial
derivatives of any function $W=W(x,y)$.

We denote by $q(u,v)$ the free charge i.e. the charge of the
collapsing scalar field, and by $\tilde{q}$ the total charge up to the
sphere of radius $r$. The latter is the charge defined by the QED
effects. The scalar field is collapsing under the influence of the
total charge $\tilde{q}$ not the free charge $q$. In a local inertial
frame: $q=\epsilon\tilde{q}$ and we define: $\tilde a \equiv
a/\epsilon$.

We write the closed system of equations
for the QED-corrected situation:

\noindent
Einstein equations:
\begin{eqnarray*}
E1  &\equiv&     r f_v   +  f g   +  {1\over 4}\alpha^2   -
                  {\alpha^2 q^2 \over 4 \epsilon  r^2} = 0 \ , \\
E2  &\equiv&     g_v  -  2 d g  +  r z^*z  =0 \ ,\\
E3  &\equiv&     d_u  -  {f g \over r^2 } -{\alpha^2 \over 4 r^2}+
                          {\alpha^2 q^2 \over 2 \epsilon  r^4}
                         +{1\over 2}(w z^*+w^* z )+{1\over 2}i e \tilde{a}(s
                         z^* - s^* z) =0 \ .
\end{eqnarray*}
Maxwell equations:
\begin{eqnarray*}
M1 &\equiv&  \tilde{a}_v - {\alpha^2q \over 2 \epsilon  r^2} = 0 \ , \\
M2  &\equiv&  q_v -i e r^2(s^* z -s z^*) =0  \ .
\end{eqnarray*}
The scalar field equations:
\begin{eqnarray*}
S1 &\equiv&  r z_u  + f z  + g w  +  i e \tilde{a} r z  + i e \tilde{a} g s +
                    {i e \over 4 \epsilon r}\alpha^2 q s = 0 \ , \\
S2  &\equiv& r w_v  +  g w  + f z  +  i e \tilde{a} r z  + i e \tilde{a} g s +
                    {i e \over 4 \epsilon  r}\alpha^2 q s = 0 \ ,
\end{eqnarray*}
and, finally, the definitions (\ref{newvar}):
\begin{eqnarray*}
D1 &\equiv& d-{\alpha_v\over\alpha}=0 \ ,  \\
D2 &\equiv& g-r_v=0 \  ,\\
D3 &\equiv& z-s_v=0 \   .
\end{eqnarray*}
The QED-corrected equations are reduced to the classical ones
\cite{HodPir} by setting  $\epsilon= 1$.
\subsection{The Characteristic Problem }
\label{char}
The system of evolutionary equations have to be completed by a
specification of the initial and the boundary data along some
characteristic hypersurface. For models involving massless fields the
characteristics are null segments. Thus, it is natural to specify the
initial conditions and the boundary conditions along null
hypersurfaces. A satisfactory, for an uncharged case, formulation of
the initial-value problem has been given by Burko \& Ori
\cite{BurOri}. The generalization for a charged case is given bellow.

We choose the initial characteristic surfaces to be: the ingoing
$v=const\equiv v_i $ hypersurface, and, the outgoing $u=const\equiv
u_i $ hypersurface. If the domain of integration includes the origin
of coordinates it leads to the necessity of a series expansion of
physical quantities in powers of the proper distance from the origin,
in a vicinity of $r=0$.  We are, however, interested in the formation
of the Cauchy horizon. We can, therefore, exclude the origin $r=0$
from the domain of integration.  We achieve this by an appropriate
choice of the final outgoing segment $u=const\equiv u_f $, so the
domain of integration does not include the origin.

Now we can remove the coordinates freedom. To do so, we fix the
``linear'' gauge, i.e. we take $r$ to be linear with $u$ or $v $ along
the characteristic hypersurfaces.  Namely, on $u=u_i $ segment we
choose $g\equiv r_v=1$, on $v=v_i $ segment we choose $f\equiv
r_u=r_{u0}$. To get $r$ along initial surfaces it is necessary to
supply $r_0=r(u_i,v_i)$ that serves as a free parameter.

The conventional choice for a characteristic segments $u_i=0$,
$v_i=r_0$, yields:
\begin{equation}
\label{r00}
r(u_i,v)=v \ ,  \   r(u,v_i)=u r_{u0} + r_0 \ .
\end{equation}
\begin{figure}[t!]
\centering
\noindent
\includegraphics[width=8cm]{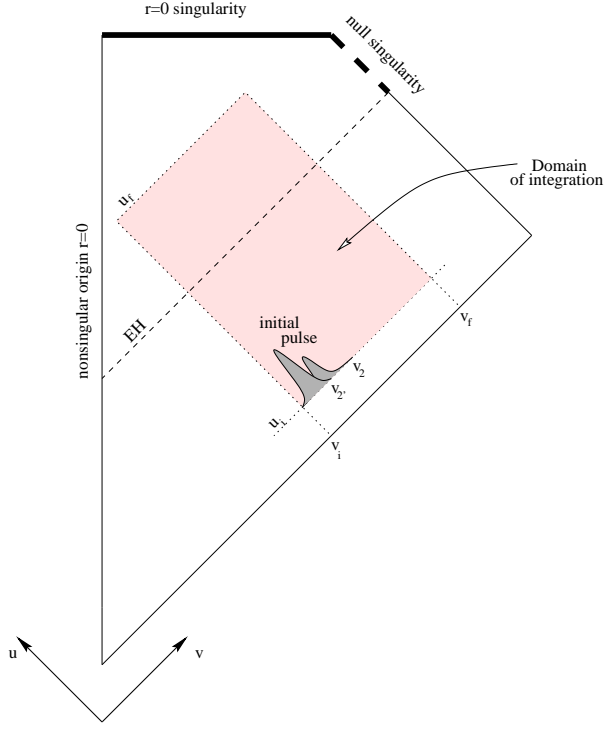}
\caption[The integrated space-time]{\label{fig:domain}
A schematic diagram of the  integrated space-time.}
\end{figure}
Now we  specify freely the scalar field distribution along the initial
segments. We choose  a compact ingoing scalar field pulse along the
ingoing $u=u_i$ segment; and the ``no-perturbation'' along the initial
outgoing $v=v_i$ segment.  Specifically, we take $\psi(u,v_i)\equiv
0$ that corresponds to a fixed static background for $v<v_i$. And we
choose $\psi(u_i,v)\equiv 0$ except at some finite region $v_1<v<v_2 ,
(v_1\ge v_i)$.  To be concrete, for a complex scalar field $\psi$,
$\psi=\phi_1+i\phi_2$, with $\phi_1, \phi_2$ two real scalar fields,
we choose:
\begin{equation}
\label{psi}
\phi_1={a \over{\sqrt{ 4 \pi}}} \sin^2\left(\pi {v-v_1 \over v_2'-v_1}\right) \ ,
\phi_2={b \over{\sqrt{ 4 \pi}}} \sin^2\left(\pi {v-v_1 \over v_2-v_1}\right) \ ,
\end{equation}
where $a, b$ are constant amplitudes, and $v_2', v_2$ are the
end-points of each of the real-fields pulses, and $v_1=v_i$ is their
common starting-point.  This choice of the initial data is
differentiable at the matching points $v_1, v_2$. The integrated
space-time is schematically depicted on Figure \ref{fig:domain}.

\noindent
From (\ref{psi}) we obtain the initial values of $z$, and $w$:
\begin{eqnarray}
\label{z}
z(u_i,v)&=&   {a\pi\over v_2'-v_1} \sin\left(2\pi {v-v_1 \over v_2'-v_1}\right)
        +     {i b\pi\over v_2-v_1} \sin\left(2\pi {v-v_1\over v_2-v_1}\right)
            \ ,\nonumber\\
w(u,v_i) & \equiv &0 \ .
\end{eqnarray}
From the constraint equations E2-E3 and the definition D1, together
with the choice $\alpha(u_i,v_i)=1$ one determines the initial values
of $d$ and $\alpha$:
\begin{eqnarray}
\label{d}
\lefteqn{d(u_i,v)=    {a^2\pi^2 v\over2 (v_2'-v_1)^2}
               \sin^2\left(2\pi {v-v_1 \over v_2'-v_1}\right) +} \nonumber\\
      & &  +   { b^2\pi^2 v\over2 (v_2-v_1)^2}
             \sin^2\left(2\pi {v-v_1\over v_2-v_1}\right)
\ .
\end{eqnarray}
\begin{equation}
\label{alf}
\alpha(u,v_i)=1 \ .
\end{equation}
We assume the Minkowski space-time for $v<v_i$, therefore, we set
$q(u,v_i)=0 $ and $ a(u,v_i)=0 $.

In our coordinates  the  mass-function (the mass-parameter) becomes:
\begin{equation}
\label{massfunc}
m(u,v)={r\over 2}\left(1+{q^2\over r^2}+{4\over\alpha^2}r_u r_v\right)
\ .
\end{equation}
Since the mass-function vanishes for the flat space-time (in the
region $v<v_i$), one can calculate: $r_{u0}=-{1\over 4 }$.  It should
be noted that for $v \gg m$ our ingoing null coordinate $v$ is closely
related (proportional) to the ingoing Eddington-Finkelstein null
coordinate $v_{\rm e}$. The $u$ coordinate is related to the proper
time of an observer at the origin. This is defined as \cite{HamSt}:
\begin{equation}
\label{properT}
T(u)=\int_0^u{\alpha(u',u')d u'}
\ .
\end{equation}
In our choice the space-time to the left to the $v=v_i$ characteristic
hypersurface is Minkowskian until the very last moments of the
collapse.  Hence, $\alpha(u,u)=1$ except the section when $u
\rightarrow u_f$, where $\alpha(u,u) \rightarrow 0$. Therefore, the
integration in (\ref{properT}) is trivial and yields: $T(u)=u$.
Later, we will find it useful to utilize the proper time of an
observer at the origin as a measure of the ``length'' of the Cauchy
horizon.
\section{The Numerical Integration Scheme}
\label{numerics}
We have converted the second order equations to  first order
equations.  Our numerical scheme is based on a simultaneous
integration of this first order system of coupled differential
equations. We solve numerically  equations $E1-E2, E4,
M1-M2, S1-S2, D1, D3, D5$. We set $\epsilon=1$ to obtain the classical
collapse with no pair creation.

The domain of integration is covered by a double-null grid. The
characteristic initial-value problem is formulated in section
\ref{char}.  The algorithm for the numerical integration in the
classical case ($\epsilon=1$) is described in \cite{HodPir}. Here we
generalize this algorithm to the case when pair creation is included.

At each step we evolve $d$ and $z$ using $E3$ and $S1$ from the
hypersurface $u$ to $u+du$. Then we solve the appropriate equations
for the rest of quantities along the outgoing null rays $u+du=const$,
starting from the initial outgoing hypersurface $v=v_i$.  We integrate
equation $D1$ to find $\alpha$, then we solve the coupled
differential equations $D2$ and $E2$ to get $r$ and $g$. Next, the
equations $D3, M2, M1$ are integrated to obtain $s$, $q$ and
$\tilde a$.  Finally, the differential equations $E1$ and $S2$ are
solved for  $f$ and $w$, respectively.  After each step in the
$u$-direction we calculate the electric field strength, $q/r^2$, along the
current outgoing null ray ($u+du =const$). We use this field value to
establish the value of $\epsilon$, according to (\ref{def_eps}), for
the next $u$-step.

This integration scheme uses three distinct methods to evolve the
initial data. All these methods are well known and commonly used (see,
for example, the recipes-book by Press W.H., et al. \cite{NumRec}).
To evolve the quantities in the $u$-direction we utilize the 5-th
order Cash-Karp Runge-Kutta method.  The differential equations in the
$v$ direction are solved using a 4-th order Runge-Kutta method.  The
integrations in the $v$ direction are performed using a three-point
Simpson method.

It is conventional to define the accuracy of a numerical method by the
scaling of the numerical error.  Thus, $n$th order accuracy means
that, the error scales as the step-size to power $n$:
\begin{equation}
\label{error}
F_{\rm real}(x)=F_{\rm calc}(x)+O(h^{n}) \ ,
\end{equation}
where $F_{\rm real}$ stands for the actual value of a function at a
point $x$, while $F_{\rm calc}$ for a calculated at the same point.
The Runge-Kutta methods, which we utilize, are all at least second
order accurate, see Ref. \cite{NumRec}.
The three-point Simpson integration method is of order $n=4$.

We have performed a few simulations with the same free parameters, but
with different grid sizes in order to check the scaling of the numerical
errors.  The numerical scheme proceeds in the $u$ and
$v$ directions on grids with corresponding grid-spacings $h_u$ and
$h_v$.  These step-sizes are connected by the numerical stability
requirement: $h_u \le f(u,v)h_v$, where $f(u,v)$ is a slowly varying
function of order unity.  We perform the convergence test by changing
the grids density in both directions.

\begin{figure}[t!]
\centering
\noindent
$(a)$\includegraphics[width=11cm]{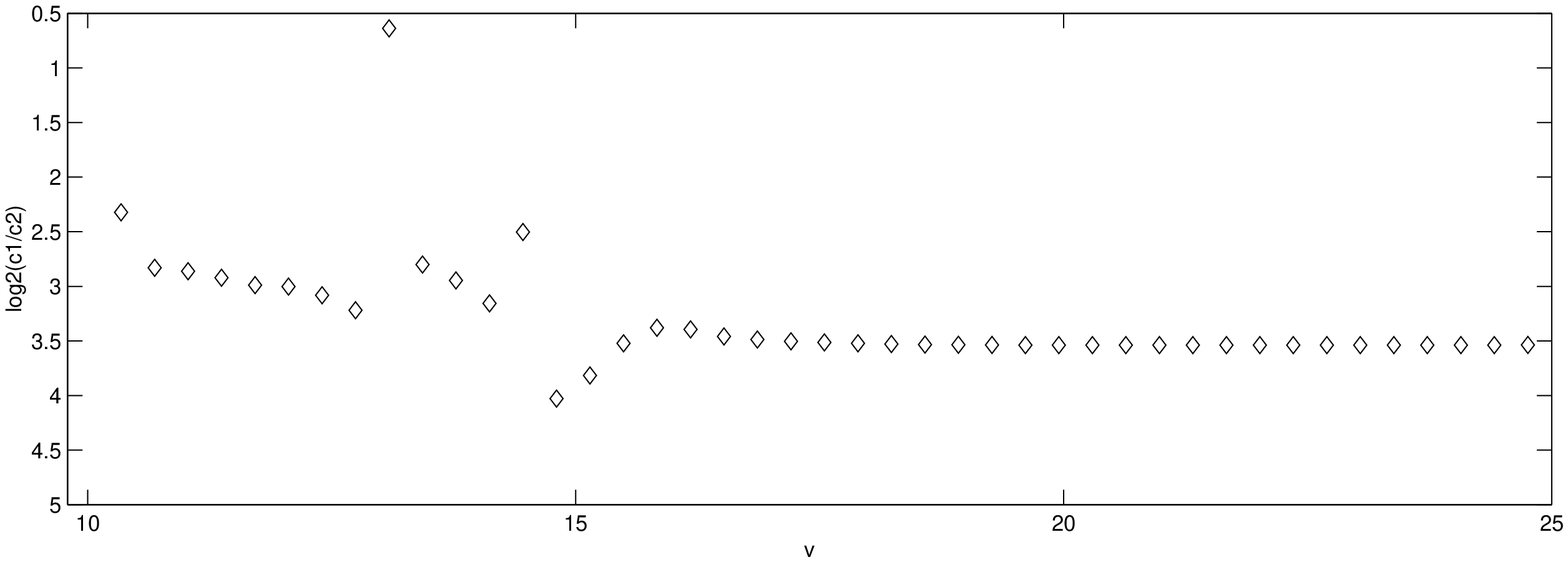}\\
\noindent
\hspace{-0.3cm}
$(b)$\hspace{0.5cm}
\includegraphics[width=10.4cm,height=9cm]{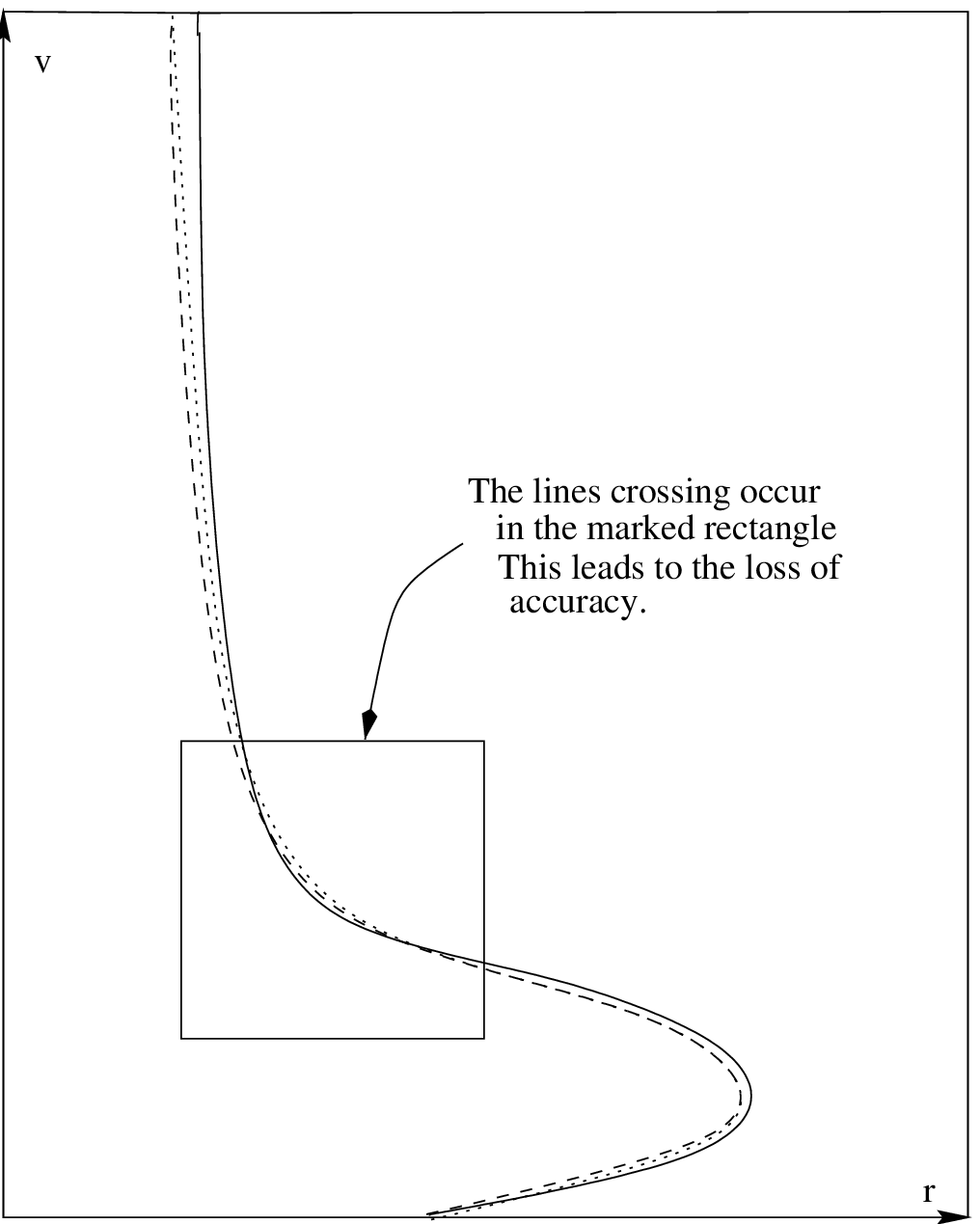}
$(c)$\includegraphics[width=11cm]{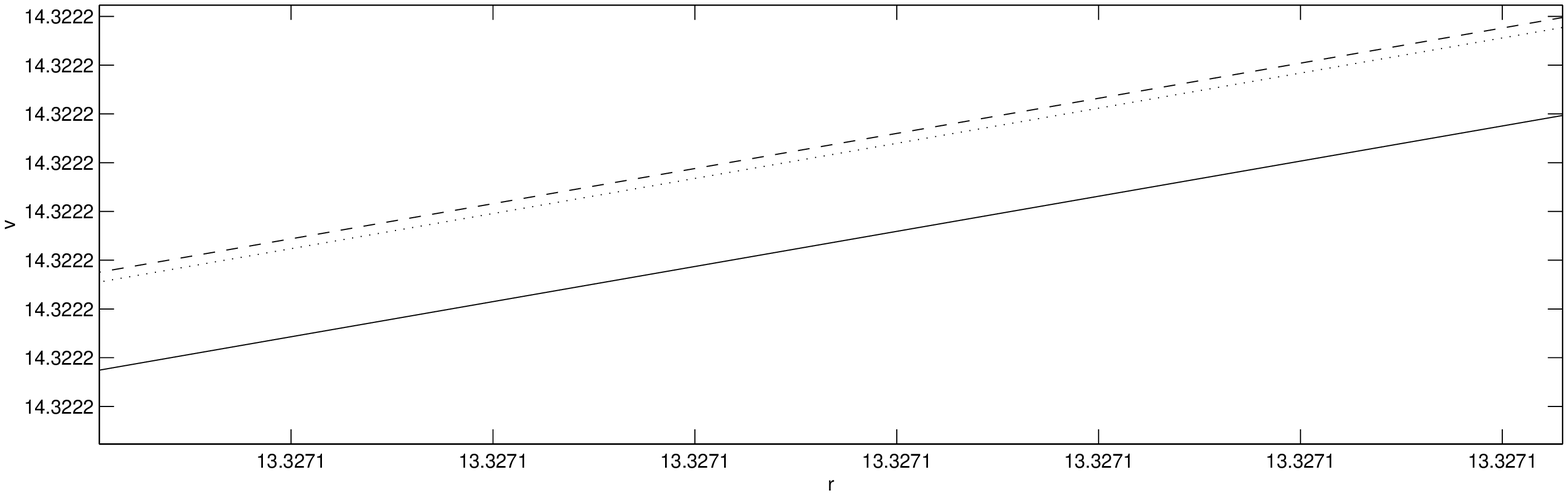}
\caption[Error scaling]{\label{fig:accur}
  $(a)$ The error scaling indicates a quadratic or higher convergence except
  at a single point with $n=0.5$.  $(b)$ To visualize the reason for
  loss of accuracy we give an approximate {\it sketch } of
  $r(u,v)$ as a function of $v$ for $u=const$, for different
  grid-densities. We give only a sketch and not the real plots, since
  in the latter the crossing is unobservable at these scales.  $(c)$
  Variation of the calculated $r(u,v)$ as a function of $v$, for
  certain value of $u$, for different grid-densities in a typical
  (with no line-crossing) region.  The grid-densities are:
  $60$ grid-points per unit interval (solid line), $120 $
  grid-points (dotted line) and $240$ grid-points (dashed line).}
\end{figure}

In what follows, $F^h$ denotes the numerically calculated, with the
numerical step-size of $h$, value of a function $F$ at some point.
Here $h$ denotes the step-size in the $v$ direction.  If the numerical
scheme, involved in a calculation of $F^h$ is convergent, then the
connection between the above quantities is given by (\ref{error}).  We
performed a series of numerical simulations with doubled grid
densities, or, equivalently, with halved step-sizes $h_v$ (and,
therefore, with halved $h_u$) .  We expect that: $F^{h} = F+O(h^{n})$.
Now, defining $c_1\equiv F^{h}-F^{h/2}$, and $c_2\equiv
F^{h/2}-F^{h/4}$, we expect that: $c_1/c_2=2^n$.  Figure
\ref{fig:accur}$(a)$ depicts $\log_2(c_1/c_2)$ along an ingoing null
$v$ coordinate, for a typical $u=const$ ray. It is clear from this
Figure that in general $n \approx 3$, indicating a third order
convergence.

One, however, notices the point $n \approx 0.5$, which indicates a very
poor convergence. Moreover, the plot looks very variable in the region
$12 \le v \le 15$. The reason for this ``jumpy'' behavior is
understood, if one looks closer at the  function $F$ itself.  We
have used  $F=r(u,v)$.  We sketch
on  Figure \ref{fig:accur}$(b)$ the radius $r(u,v)$ along an
$u=const$ outgoing null ray (see the next section). This sketch is
magnified, since the actual plots of $r$ for different grid
densities are indistinguishable on this scale.  In the marked box one
observes the crossing of curves for different grid densities. This crossing
leads to a decrease of the convergence order. Notably,
after this crossing the convergence returns to the high order.

Figure \ref{fig:accur}$(c)$ displays the variation of the calculated
$F$ as a function of $v$ along a typical $u=const$ ray for different
grids densities.  The observed picture confirms the convergence.
\section{ Results}
\label{results}
\subsection{Classical Charged Collapse}
\label{ClassRes}
We begin with verifying our numerical code for a classical collapse, a
collapse without quantum effects. We set $\epsilon=1$.  We fix the
free numerical parameters to define the problem: $a=0.5, b=0.461,
v_i=r_0=10, v_f=90, v_2=16, v_2'=20, e=0.15$. The number of
grid-points along outgoing and ingoing rays is of order of $10^2$ per
unit interval.  We follow the evolution of the regular initial data
via the formation of an apparent horizon and a Cauchy horizon, toward
the central singularity.  For this specific choice of free parameters
the resulting black hole has a charge-to-mass ratio, $q/m \approx
0.98$, in geometrical units.

\begin{figure}[h!]
\centering
\noindent
\includegraphics[width=9cm]{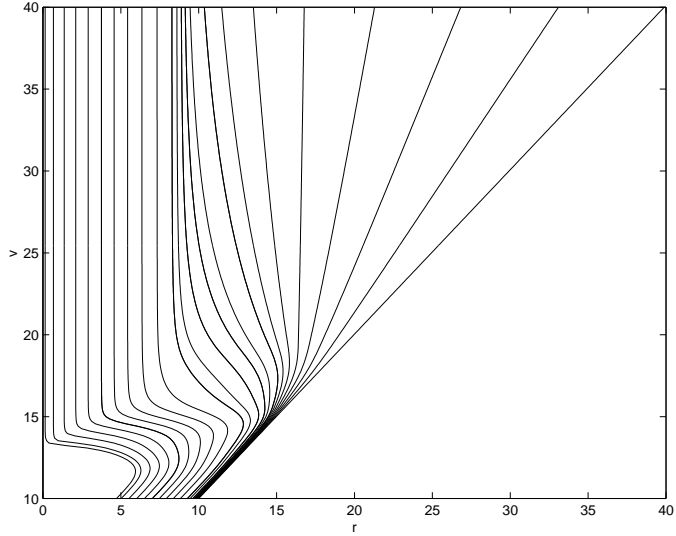}
\caption[Radius $r(u,v)$ vs. $v$]{\label{fig:r_v}
 Radius $r(u,v)$ as a function of
  $v$ along  $u=const$ rays. The retarded time $u$ increases
  from the initial straight ray toward the origin (from the right to
  the left), indicating the course of evolution of the space-time.} 
\end{figure}

Figure \ref{fig:r_v} displays the metric function $r(u,v)$ as a
function of the ingoing coordinate $v$ for different values of the
outgoing null coordinate $u$. One can distinguish between two
 kinds of rays: the outer and the inner rays. The outer rays,
outgoing rays at small $u$, escape to infinity: $
r(u,v\rightarrow\infty) \longrightarrow\infty$. Increasing $u$, one
finds the first outgoing ray, which does not escape to infinity, but
tends to a constant value.  This ray indicates the event horizon,
which asymptotically coincides with the apparent horizon. The ray
becomes vertical at about $r\approx 16$ (not displayed).  Increasing
$u$ further, one steps into the inner region of the integrated
space-time. Outgoing rays in this region approach a constant radius,
depending on $u$, in asymptoticly late advanced times
$v\rightarrow\infty$. This indicates the formation of a Cauchy horizon.
Following the evolution of the space-time, the solution approaches the
origin, $r=0$.  Since our code is not constructed to include the
origin we stop at $u_f$, before the integration reaches the origin.

The main feature seen here is the existence of
a contracting Cauchy horizon. The Cauchy horizon  is not a stationary $r_{\rm CH}=const$
hypersurface as in the case of a classical Reissner-Nordstr{\o}m space-time, but it depends
on the outgoing null coordinate $u$, namely, it contracts toward the
origin $r=0$ in late retarded  time $u$.
\begin{figure}[h!]
\centering
\noindent
\includegraphics[width=9cm]{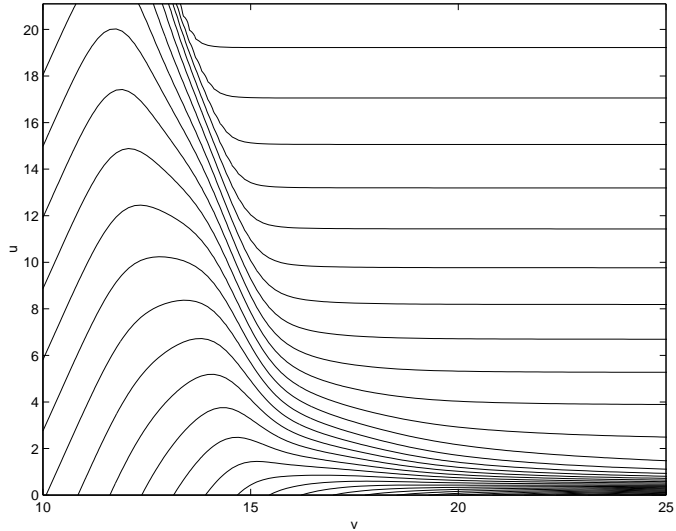}
\caption[Contour lines of the radius $r(u,v)$ in the $u v$ plane]{\label{fig:r_uv}
 Lines of constant radius in $u v$-plane. The radius
  decreases from the bottom toward the top of the Figure.
  The apparent horizon,  along which $r_v=0$, separates
  the exterior and the interior of the black hole. The latter region contains
  the Cauchy horizon. }
\end{figure} 
\begin{figure}[t!]
\centering
\noindent
\includegraphics[width=9cm]{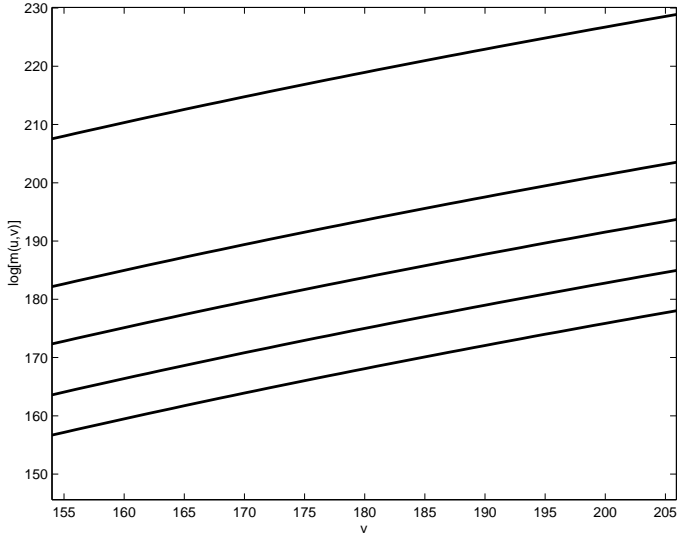}
\caption[Mass-function $m(u,v)$ vs. $v$]{\label{fig:mass1}
 Logarithm of the mass-function, $m(u,v)$, vs. $v$ along $u=const$
  null rays.  The linear dependence of $m(u,v)$ on late time $v$
  indicates the mass-inflation.  The retarded time $u$ increases
  toward the top of the Figure.} 
\end{figure}

Figure \ref{fig:r_uv} depicts constant radii contour lines in the $u
v$-plane. The bottom of the Figure ($u=0$) is the initial regular
hypersurface. Looking along the  $u$ direction one soon observes the formation
of an apparent horizon, along which $r_v=0$. The apparent horizon
separates the two regions with $r_v>0 $ (on the left), and $r_v<0$ (on the
right). The latter contains, also, the asymptotically
($v\rightarrow\infty$) constant, $u$ dependent section $r_{\rm
  CH}=const(u)$, representing the Cauchy horizon.  The Cauchy horizon \ itself is a null
hypersurface which is approached as $v$ goes to infinity.

We depict the logarithm of the mass-function (\ref{massfunc}) in
Figure \ref{fig:mass1}, for a sequence of $u=const$ null rays.  The
straight lines indicate the exponential dependence of the
mass-function on $v$ for late advanced times. The exponential growth
confirms the conclusion that a mass-inflation  indeed takes
place in this collapse.

We have performed simulations with different free parameters: changing
the amplitudes $a$ and $b$, the elementary charge $e$, stretching and
squeezing the domain of integration. In all these cases we have not
seen any qualitative difference between the results.  The results,
which we have obtained, are in excellent agreement with the previously
established results, and fit well those in \cite{HodPir}.  We
conclude, thus, that our numerical code gives correct results to this
case.  We turn now to the problem of collapse with pair creation.
\subsection{Collapse with Pair Creation }
\label{QEDRes}
The critical electric field, $E_{\rm cr}$, is an additional free parameter in
the  problem with pair creation.  The critical field strength $E_{\rm cr}$ is chosen
in the forthcoming graphs so that it is reached just after the
formation of the apparent horizon.  This choice is arbitrary. Other
comparable values of $E_{\rm cr}$  lead to a qualitatively similar results,
unless $E_{\rm cr}$ is reached long before the formation of the apparent
horizon (see bellow).

\begin{figure}[t!]
\centering
\noindent
\includegraphics[width=9cm]{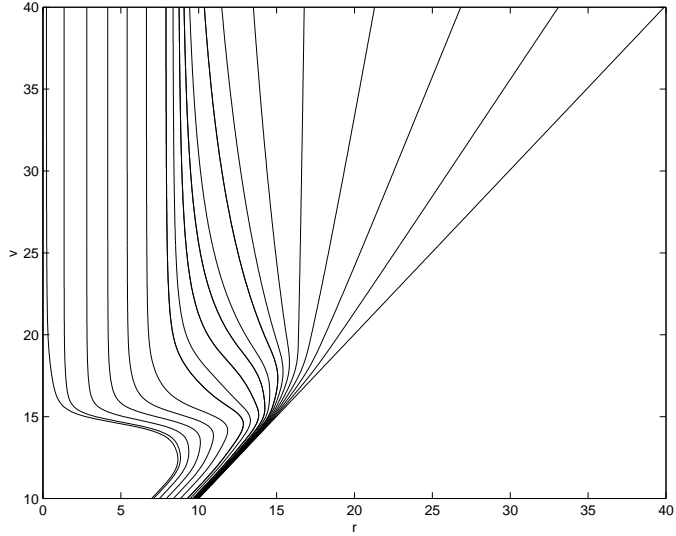}
\caption[Radius vs. $v$ with QED-corrections]{\label{fig:r_v_qed}
  The radius, $r(u,v)$, as a function of $v$ along $u=const$ rays for
  collapse with pair creation.  The retarded time $u$ increases to the
  left. One still observes the formation of the Cauchy horizon \.
  There is also an apparent incline of the late retarded rays toward
  the origin, indicating discharge and signaling the destruction of
  the Cauchy horizon. }
\end{figure}

Figure \ref{fig:r_v_qed} displays the radius $r(u,v)$ as a function of
$v$ for a sequence of $u=const$ null rays. This Figure is analogous to
Figure \ref{fig:r_v} for the classical situation.  On a first sight
these Figures seem very similar. There are, however, obvious
differences, both qualitative and quantitative. A closer look uncovers
the different incline of the late retarded rays ($u \rightarrow u_f$):
while these rays depicted on Fig. \ref{fig:r_v} are practically
vertical, those on Fig. \ref{fig:r_v_qed} have an observable incline
toward the origin $r=0$. The very last ray on the latter Figure, has
an apparent tendency toward the origin, indicating that the $r=0$
strong singularity is close.  Other signs of the difference between
the two situations are quantitative ones.  The whole ``life-time'' (in
terms of the retarded $u$-time) of the QED-corrected system, before it
hits the $r=0$ singularity is significantly shorter than the
``life-time'' of the corresponding classical system. For example, for
the set of free parameters (see at the beginning of the previous
section) the evolution of the QED-corrected system lasts $u_f\approx
12$, while for a classical system it takes about $u_f\approx 21$,
before it crushes into the $r=0$ singularity. We note that $u_f$ is
the proper time of an observer at the origin (\ref{properT}).
\begin{figure}[h!]
\centering
\noindent
\includegraphics[width=9cm]{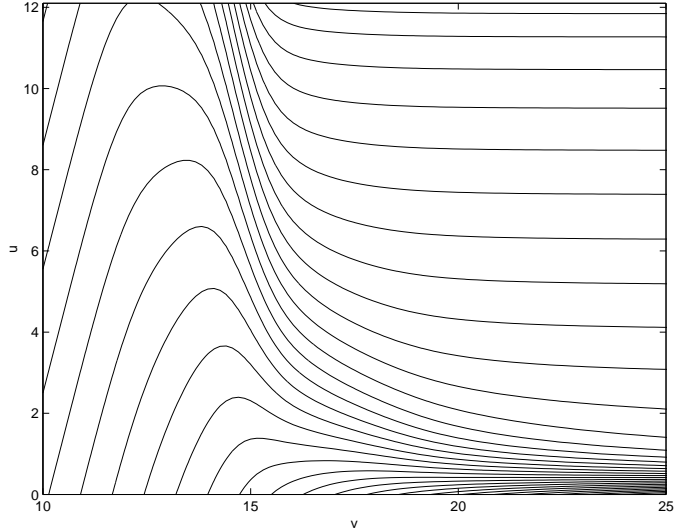}
\caption[Contour lines of radius $r(u,v)$ in $u v$-plane,
 with QED-corrections]{\label{fig:r_uv_qed}
Contour lines of constant radius in $u v$-plane
  for a collapse with pair creation. The radius decreases from the
  bottom toward the top of the Figure. One observes the apparent
  horizon and the Cauchy horizon .}
\end{figure}

\begin{figure}[t!]
\centering
\noindent
\includegraphics[width=9cm,height=8cm]{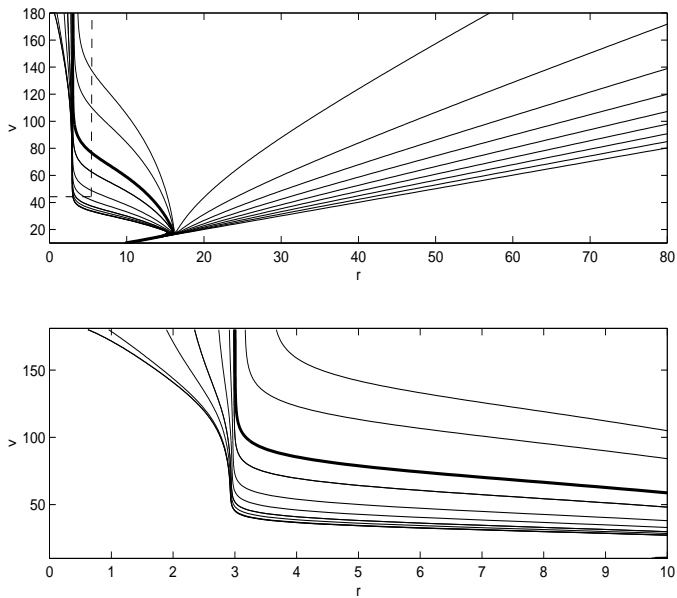}\\
\caption[Discharge of the black hole with $Q/M \ll 1$]
{\label{fig:SmallQM} The radius $r(u,v)$ vs. $v$ along a sequence of
  $u=const$ rays in the situation when the charge-to-mass ratio, seen
  from infinity, of the formed black hole is small compared to unity.
  The bottom panel displays an enlargement of the marked region on the
  top panel.  The rays' incline toward the origin is apparent,
  indicating the discharge. The heavy line marks the beginning of the
  discharge.  }
\end{figure}

Figure \ref{fig:r_uv_qed} displays the contour lines of constant radii
in the $uv$-plane.  This Figure is analogous to Figure \ref{fig:r_uv} for
the classical situation.  Again one can observe a different incline of
the contour lines in the late retarded and advanced time regions.
This difference in the incline between the classical and the
QED-corrected problems can be interpreted as just a close approach to
the intersection of the Cauchy horizon  with the strong space-like $r=0$
singularity during the numerical simulation. The straightening of the 
outgoing rays occurs later in terms of the advanced $v$-time: the
curvature is high in the vicinity of the strong singularity turning
the rays toward the singular origin.
The discharge is more apparent in the situation when the
black hole, as it seen from infinity, has the small charge-to-mass
ratio, see Figure \ref{fig:SmallQM}.

We define the ``length'' of the Cauchy horizon as a proper time of an
observer at the origin between the moment when he or she emits a
last outgouing light signal that escapes to infinity (the event horizon) and
the moment  when he or she emits a first outgoing light signal that
unavoidably falls into the spacelike singularity. This ``length'' is
related to the retarded time $u$, see  (\ref{properT}). We observe,
therefore, that in the collapse with discharge the ``length'' of 
the Cauchy horizon  is ``shorter'', compared to the classical case.
\begin{figure}[h!]
\centering
\noindent
\includegraphics[width=13cm,height=15cm]{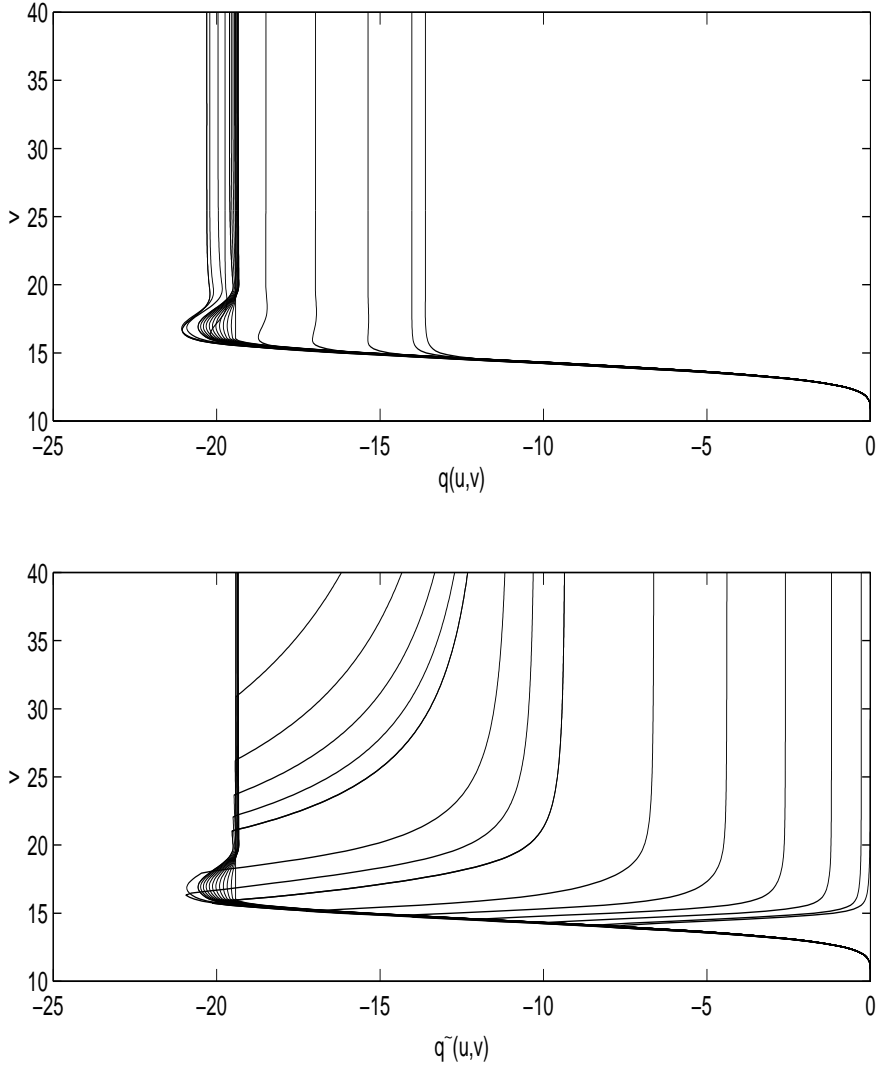}
\caption[The free and the total charge in the integrated space-time]
{\label{fig:q_q}
 Top panel: the free charge $q(u,v)$ along $u=const$ null rays.
  Bottom panel: the total charge ${\tilde q}(u,v)$ along the same
  $u=const$ null rays. On both panels the retarded $u$ time increases from
  the leftmost ray toward the rightmost ray.}
\end{figure} 
\begin{figure}[t!]
\centering
\noindent
\includegraphics[width=9cm]{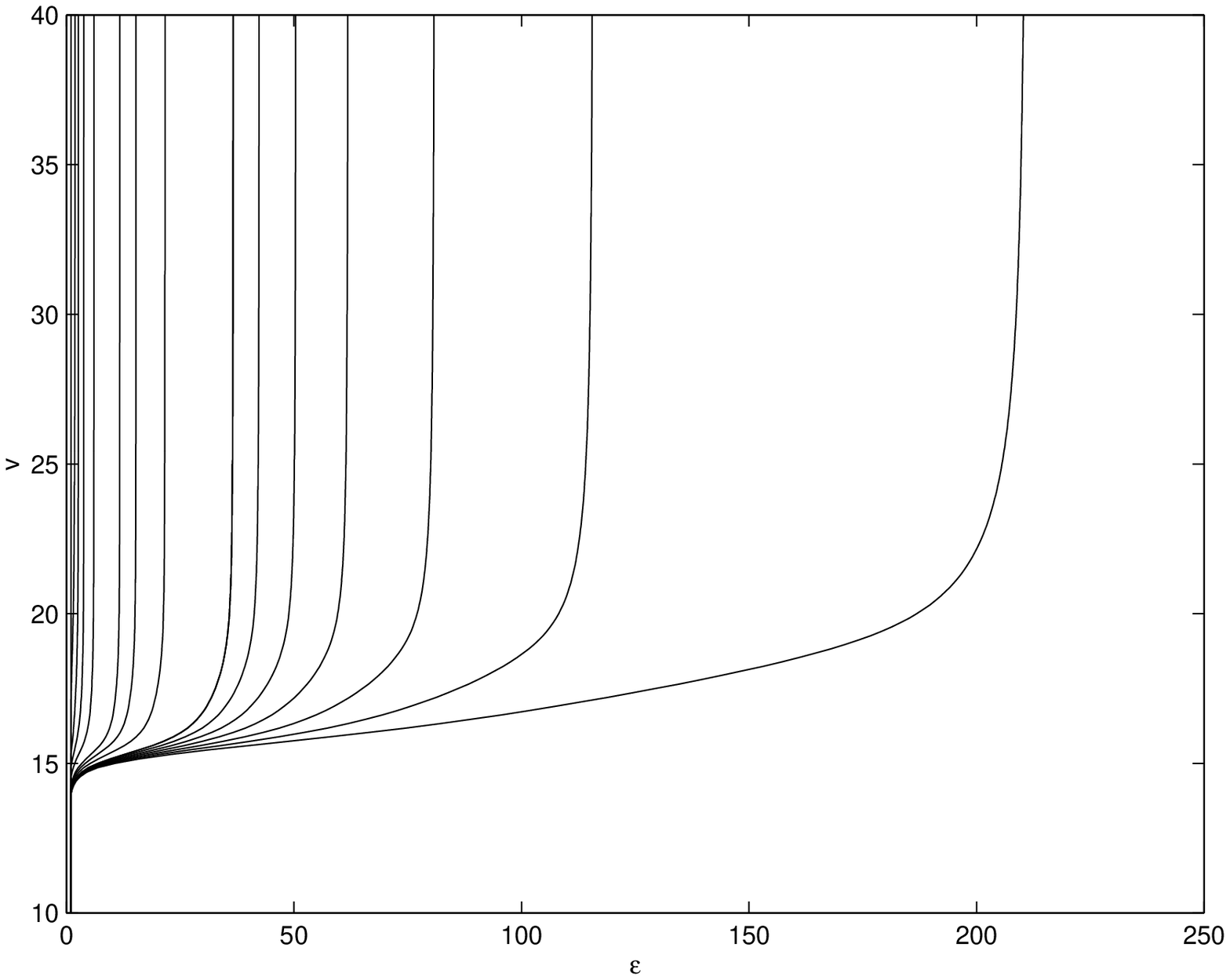}
\caption[The dielectric constant $\epsilon(u,v) $ vs. $v$]{\label{fig:epsilon}
  The dielectric constant $\epsilon(u,v) $ vs.$ v$ along $u=const$
  null rays. The retarded $u$ time increases from the left to the
  right.  The dielectric constant, which represents the
  ``polarization'' field, grows with time $u$.}
\end{figure}

On the top panel of Figure \ref{fig:q_q} we depict the free charge
$q(u,v)$ as a function of the advanced time $v$ along $u=const$ rays.
The bottom panel of Figure \ref{fig:q_q} displays the total,
QED-corrected, charge ${\tilde q}(u,v)$ along the same $u=const$
outgoing surfaces. The common characteristic property of the graphs is
the straightening of the $u=const $ rays in late advanced times, when
approaching the formed Cauchy horizon. The $u=const$ rays intersect
the Cauchy horizon \ as $v\rightarrow\infty$ and for different $u$
values the intersection occurs in different points with $r_{\rm
  CH}=const(u)$.  Hence, a charge measured along an outgoing null rays
approaches a constant $u$-dependent value which decreases with $u$,
indicating a contraction of the Cauchy horizon \ to the $r=0$
singularity.  The difference between the graphs is the strong decay of
the charge in the QED-corrected case relative to the classical case.
In the former one, the charge approaches a maximum, which is defined
by the strength of the critical field $E_{\rm cr}$, and then it is
reduced by the created pairs keeping the electric field at constant
$E_{\rm cr}$ value.  After the pair creation process brought the
charge to the ``right'' value, according to (\ref{def_eps}), the
charge remains constant along an outgoing $u=const$ rays going toward
the Cauchy horizon.  Moreover, one observes the decrease to nearly
zero charge on the bottom panel, in contrast to the classical
situation.

\begin{figure}[t!]
\centering
\noindent
\includegraphics[width=9cm]{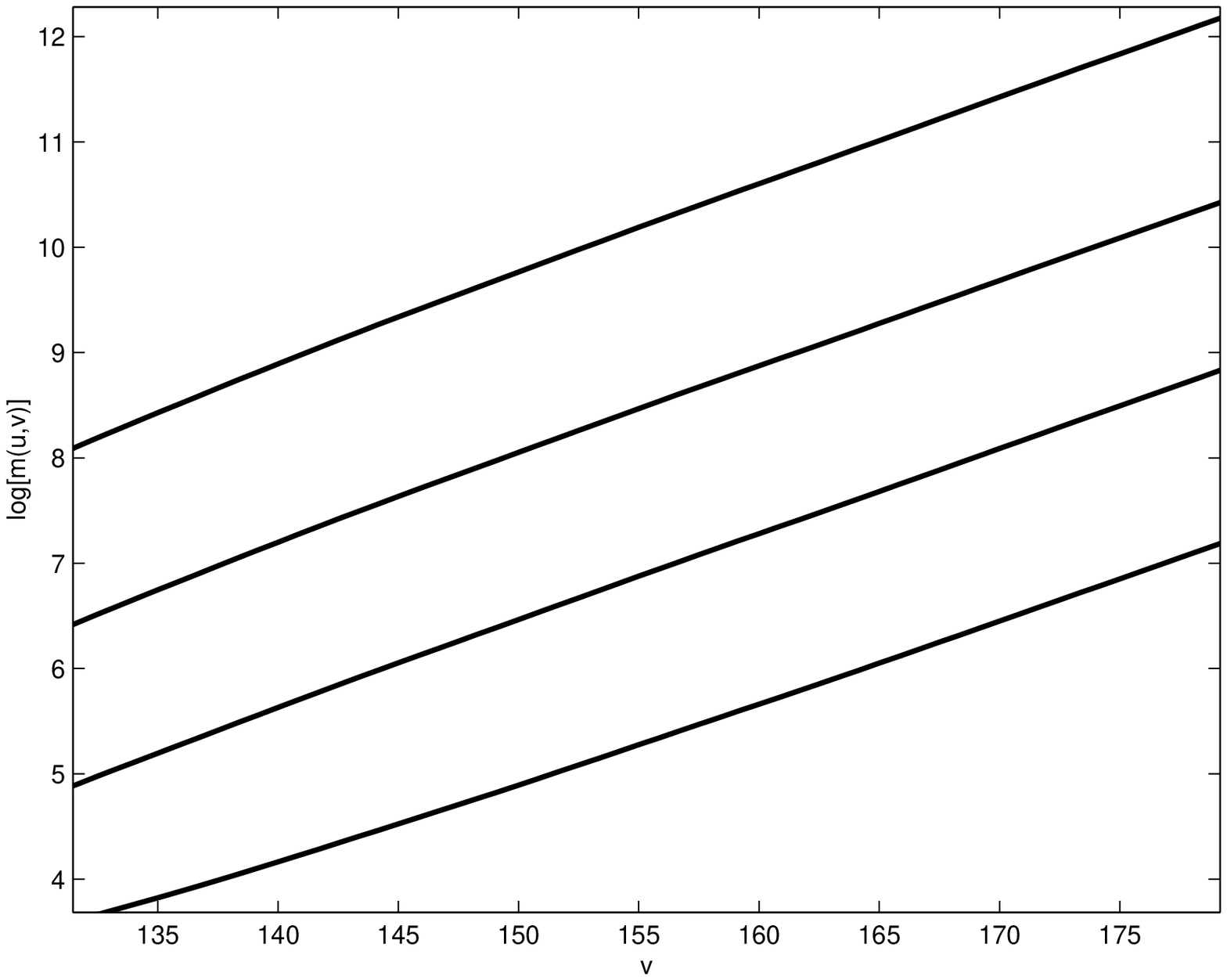}
\caption[The mass-function $m(u,v)$ vs. $v$, with QED-corrections]
{\label{fig:massinfQED}
 The logarithm of the mass-function, $m(u,v)$ vs. $v$ along a
  sequence of $u=const$ null rays.  Again, the linear dependence of
  $m(u,v)$ on the late time $v$ indicates the mass-inflation. The
  retarded time $u$ grows from the bottom to the top.}
\end{figure}  
The effective dielectric constant $\epsilon$ is depicted as a function
of the advanced time $v$ in Figure \ref{fig:epsilon}.  We note the
growth of $\epsilon(u,v)$ with the lapse of the retarded time $u$ (the
rightmost ray in Figure is the late $u=const$ depicted ray, the
leftmost ray is the early $u=const$ ray). This is readily understood:
the electric displacement ${\bf D}$ is growing and in order to keep
the electric field at the constant $E_{\rm cr}$ value, the polarization
field, represented by $\epsilon$, must grow, also.

In Figure \ref{fig:massinfQED} we depict the logarithm of the
mass-function $m(u,v)$ along a sequence of $u=const$ outgoing null
rays. The exponential dependence is clear, indicating that a
mass-inflation takes place also in the collapse with the pair
creation.

\begin{figure}[t!]
\centering
\noindent
$(a)$\includegraphics[width=9cm,height=6.1cm]{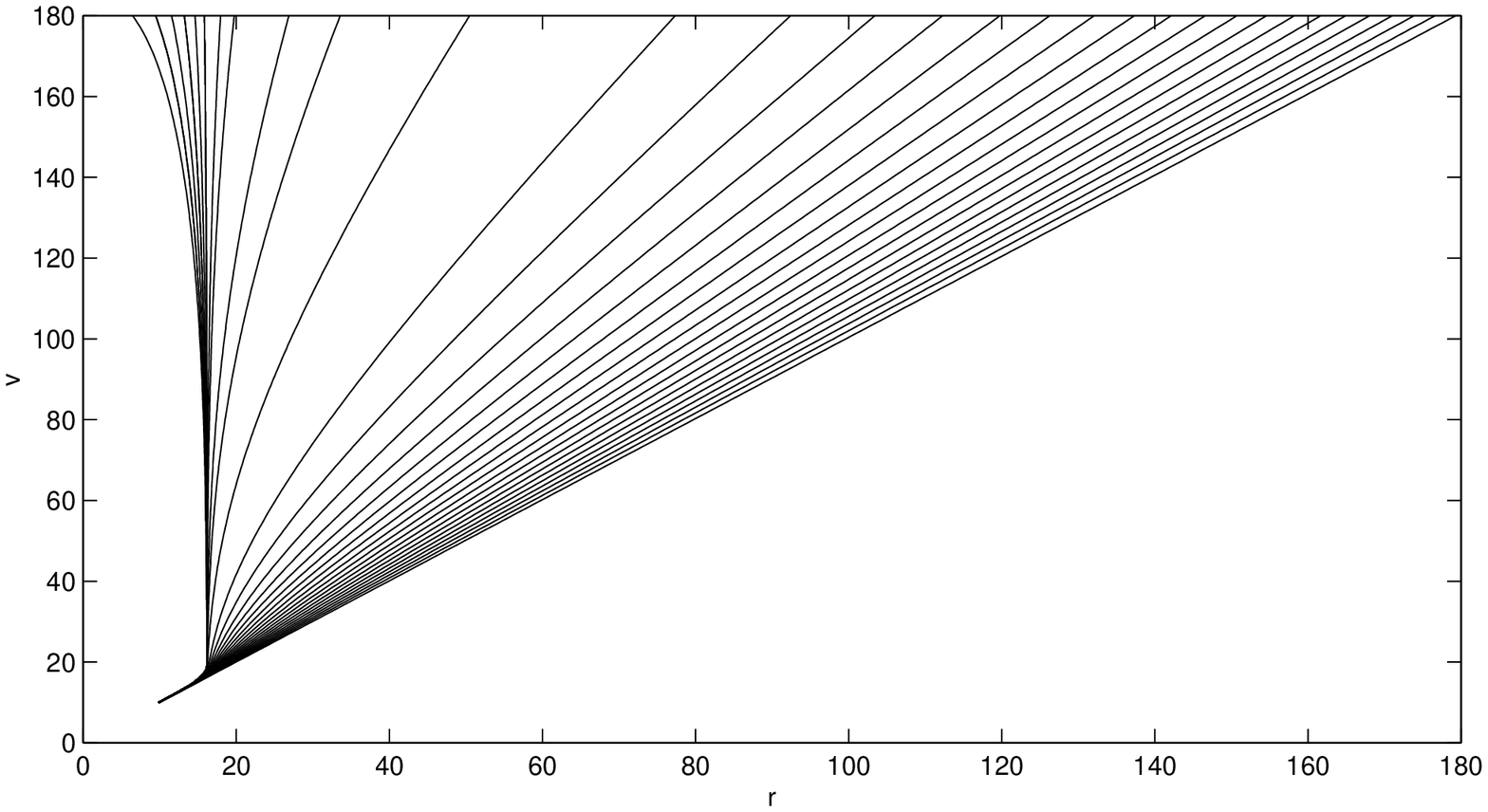}\\
\noindent
\hspace{-0.1cm}
$(b)$\hspace{-0.1cm}
\includegraphics[width=9cm,height=6.1cm]{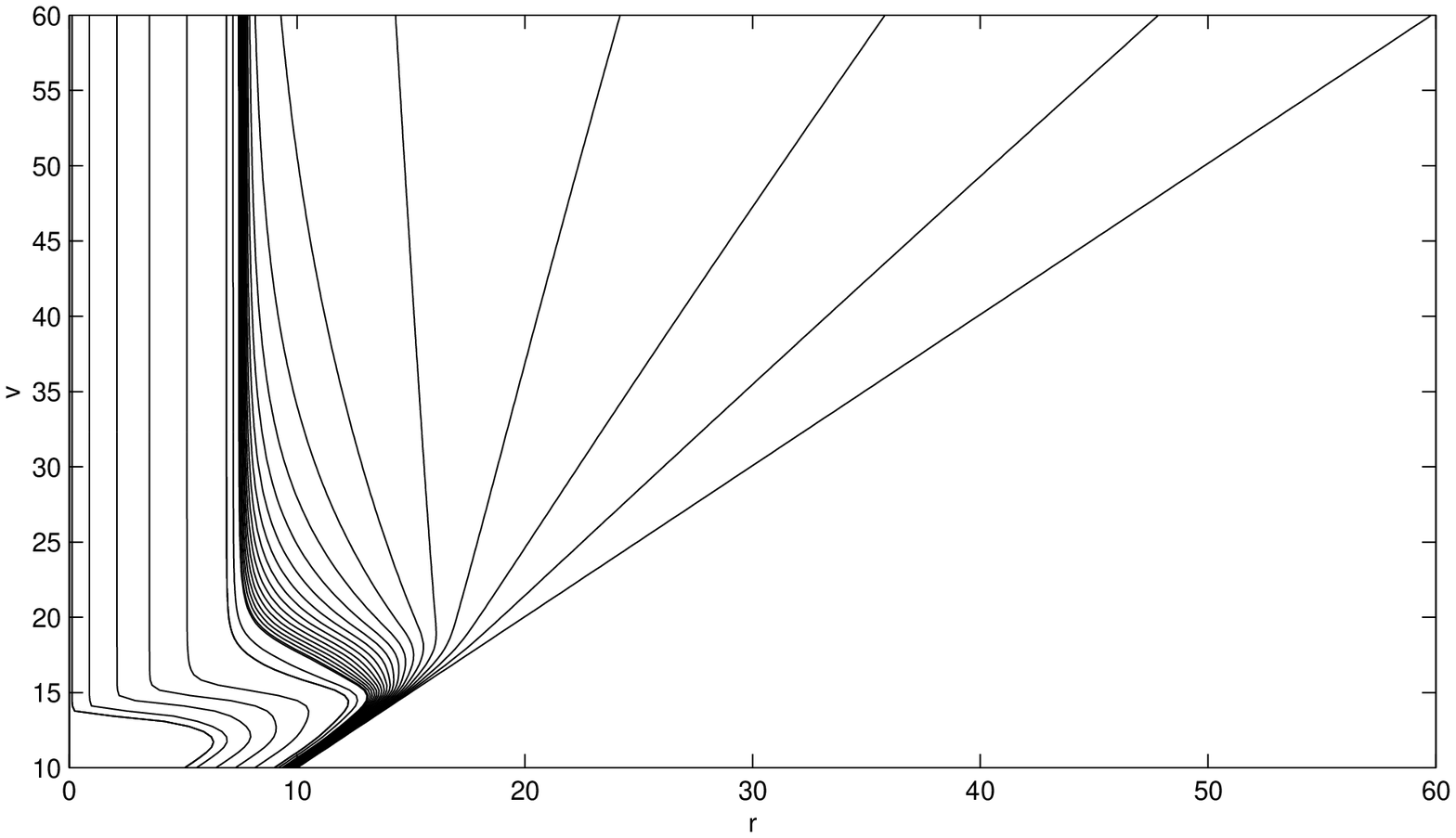}
\caption[The early and the late discharge]
{\label{toextrims}  The radius
  $r(u,v)$ vs.$v$  along a sequence of $u=const$ rays in two
  situations: (a) The discharge begins at an  early moment of
  collapse - the Cauchy horizon  is unobservable and late rays are falling to
  the origin; (b) The discharge begins deep
  inside the apparent horizon of the  black hole. The Cauchy horizon  is
  almost unaffected relative the corresponding classical situation.}
\end{figure}

In the simulations presented so far $E_{\rm cr}$ was chosen such that
$E_{\rm cr}$ was reached just after the apparent horizon forms.  We
have also performed simulations in which the critical field was
reached $(i)$ long before or $(ii)$ long after the apparent horizon
formed.  Figure \ref{toextrims}(a) displays the space-time of the
black hole which is formed in the situation $(i)$. In this case the
discharge begins long before the black hole forms. The charge of the
black hole is very small, $q \ll m$.  Therefore $r_- \approx {Q^2
  \over M c^2} \ll r_+$ and the Cauchy horizon  is unobservable, i.e.
it forms (if at all) in the domain of very high (Planck) curvature.
Figure \ref{toextrims}(b) displays the space-time of a black hole
which is formed in the situation $(ii)$. In this case the picture is
very similar to the classical one. The pair creation begins at a very
late stage when a significant part of the Cauchy horizon  is already
formed. The created pairs affect only the last stages of the evolution
and the full formed Cauchy horizon  is only slightly shorter than in
the classical situation.
\newpage
\section{Summary, Conclusions and Outlook}
\label{Summary}
We have constructed a dynamical model of a collapse of
self-gravitating electrically charged massless scalar field including
pair creation in strong electric fields.

Previous studies of the problem were concerned with particles
production in the electric field: $(i)$ outside
\cite{ZCGDR1,ZCGDR2,ZCGDR3,ZCGDR4} or $(ii)$ inside
\cite{NovStar,HerHis} the event horizon of a pre-existing charged
black hole.  Here, we have formulated the problem in a way allowing us
to address the question of the influence of the QED-effects on a
formation of black holes within the framework of an {\it evolutionary
  model}. Our particular interest was devoted to a dynamical formation
of the Cauchy horizon.

We have presented and used a toy model that treats the effect of pair
creation in strong electric fields as an appearance of a local
effective dielectric constant.  The characteristic field strength
$E_{\rm cr}$ describing the quantum effects in external field is viewed as a
boundary between the classical and the quantum stages of the system's
evolution.  We have simulated the collapsing matter by a massless
scalar field. Hence, this critical field $E_{\rm cr}$ was set as a free
parameter which defined the mass of the created particles according to
(\ref{CriticalField}).  The conclusions from the numerical integration
are presented below.

If the critical electric field strength is reached long before the
formation of an apparent horizon almost complete discharge of the
collapsing matter takes place and the final black hole is an almost
neutral Schwarzschild black hole, see Figure \ref{toextrims}(a).  The
black hole's charge, as is measured from infinity is not strictly
zero. The remaining charge is of order $E_{\rm cr} M^2$, where $M$ is the
mass of the black hole.

If the critical value $E_{\rm cr}$ is approached after the apparent
horizon forms the black hole seems from outside to have all its
initial charge.  Nevertheless, the process of discharge takes place in
the inner region of a black hole. In the classical collapse strong
central Schwarzschild-like singularity, and a weak singular Cauchy
horizon is formed.  With pair creation only a fraction of the Cauchy
horizon remains a weak, null singular. The rest is replaced by a
strong spacelike singularity. The ``length'' (as defined in
section (\ref{QEDRes}) of this weak singular section
of the Cauchy horizon depends on the critical field strength $E_{\rm
  cr}$.

It is interesting to consider the interplay between the pair creation
effects considered here and Planck scale quantum-gravity physics.
There are two ``extreme'' situations: $(i)$ The critical field is
reached deep inside the inner region, in the Planck region surrounding
the central singularity. Then the Cauchy horizon  is unaffected (or,
it is unclear how the horizon is affected at the Planck scales) by the
pair creation effect. It remains just the same as in a classical
collapse (see Fig.  \ref{toextrims}(b)); $(ii)$ The critical field is
reached at some moment before the formation of the  apparent horizon
then the Cauchy horizon is not formed at all (see Fig.
\ref{toextrims}(a)).  More precisely, as it seen by an infalling
observer, the formation of the Cauchy horizon occurs on the Planck
scale. That is, from a practical point of view the Cauchy horizon does
not form, cf. \cite{NovStar}.  Between two extremes $(i)-(ii)$ the
``length'' of a weak singular, null section of a Cauchy horizon varies
between a full, classical ``length'' and zero.

It should be emphasized that even if the critical field is approached
before the formation of an apparent horizon , but after the moment
$(ii)$, the weak, null singular section of a Cauchy horizon survives.
This section is the familiar null singularity, and an internal
mass-parameter diverges exponentially approaching it (see Fig.
\ref{fig:massinfQED}). We observe in the simulations the formation (or
the lack) of a Cauchy horizon  in various situation, mentioned
before.

A charged spherically symmetric collapse is not a generic phenomena in
the nature.  One does not expect, in general, a significant excess of
charge. But even if this not a case, the initially charged
self-gravitating matter distribution will be rapidly neutralized by an
accretion of an inter-stellar matter and by the pairs creation
process.

A more generic phenomena of a collapse is one endowed by an angular
momentum. This is an axisymmetric process and it is rather
difficult to study, both analytically and numerically.  The common
property of a charged spherically symmetric collapse and the general
axisymmetric one is an existence of Cauchy horizons inside the  black
hole. Hence, we use the charged collapse as a toy model to simulate
the more general rotating situation and to derive a conclusions about
the inner structure of the spinning black holes.

It would be interesting question to apply the considerations of this
work to a local production of particles below the event horizon of a
rotating black hole.  Outside the event horizon of a rotating black
hole the coupling of the black hole's spin to the orbital angular
momentum of particles leads to the phenomena of superradiance. If this
process takes place below the event horizon of a spinning black hole,
then this may lead to a destruction of a Cauchy horizon, just like in
a charged case.

\noindent
{\bf ACKNOWLEDGMENTS} We thank S. Hod and S. Ayal for helpful
discussions. The research was supported by a grant from the Israel
Basic Research Foundation.

\end{document}